\documentclass[submission,copyright,creativecommons]{eptcs}
\providecommand{\event}{ICLP 2021}

\usepackage{mathptmx}
\usepackage{xcolor}
\usepackage{url}
\usepackage{amsmath}
\usepackage{amsfonts}
\usepackage{hyperref}

\newtheorem{example}{Example}
\newtheorem{theorem}{Theorem}
\newtheorem{proof}{Proof}

\title{Syntactic Requirements for Well-defined Hybrid Probabilistic Logic Programs}
\author{Damiano Azzolini 
\institute{University of Ferrara \\ Dipartimento di Ingegneria} 
\email{damiano.azzolini@unife.it}
\and 
Fabrizio Riguzzi
\institute{University of Ferrara \\ Dipartimento di Matematica e Informatica} 
\email{fabrizio.riguzzi@unife.it}
}
\def\titlerunning{Syntactic Requirements for Well-defined Hybrid Probabilistic Logic Programs}
\def\authorrunning{D. Azzolini \& F. Riguzzi}

\begin{document}


\maketitle

\begin{abstract}
Hybrid probabilistic logic programs can represent several scenarios thanks to the expressivity of Logic Programming extended with facts representing discrete and continuous distributions.
The semantics for this type of programs is crucial since it ensures that a probability can be assigned to every query.
Here, following one recent semantics proposal, we illustrate a concrete syntax, and we analyse the syntactic requirements needed to preserve the well-definedness.
\end{abstract}


\section{Introduction}
\label{sec:introduction}
The power and expressivity of Probabilistic Logic Programming (PLP)~\cite{DBLP:journals/ml/RaedtK15,Rig18-BKaddress} have been utilized to represent many real world situations~\cite{azzolini2019studying,DBLP:conf/ijcai/RaedtKT07,NguRig17-IMAKE-BC}.
Usually, probabilistic logic programs involve only discrete random variables with Bernoulli or Categorical distributions.
Numerous solutions emerged to also handle continuous distributions~\cite{DBLP:conf/ilp/GutmannJR10,DBLP:journals/ai/MichelsHLV15,DBLP:journals/corr/abs-1807-00614}, increasing the expressiveness of PLP and giving birth to \textit{hybrid} probabilistic logic programs, that is, programs that include discrete and continuous random variables.
Inference in this type of programs is hard since it combines the complexity of the grounding computation with the intractability of a distribution defined by a mixture of random variables. 

Usually, inference in general hybrid probabilistic logic programs (i.e., without imposing restrictions on the type of distributions allowed) is done by leveraging knowledge compilation and using external solvers~\cite{DBLP:journals/corr/abs-1807-00614} or by sampling~\cite{AzzRigLamMas19-AIXIA-IC,Rig13-FI-IJ}.

The semantics of hybrid programs has been proved well-defined in~\cite{AzzRigLam21-AIJ-IJ}, but the authors neither provided an explicit syntax, nor introduced explicit syntactic requirements. 
In this paper, we close this gap by introducing an explicit syntax and the syntactic requirements needed to preserve the well-definedness of the semantics of hybrid probabilistic logic programs.

The paper is structured as follows: in Section~\ref{sec:plp} we introduce the semantics and the syntax for programs with only discrete random variables with both finite and infinite number of explanations.
The semantics and syntax for hybrid programs is introduced in Section~\ref{sec:hybrid_plp}, and the syntactic requirements are analysed in Section~\ref{sec:syntax_req}.
Section~\ref{sec:conclusions} concludes the paper.
All the Sections are accompanied by several examples, to make the concepts clearer. 

\section{Probabilistic Logic Programming}
\label{sec:plp}
Probabilistic Logic Programming (PLP) allows the construction of complex yet interpretable probability models through the usage of Logic Programming.
Several probabilistic logic languages based on the Distribution Semantics (DS)~\cite{DBLP:conf/iclp/Sato95} have been proposed during the years, such as PRISM~\cite{DBLP:conf/iclp/Sato95}, and ProbLog~\cite{DBLP:conf/ijcai/RaedtKT07}. 
Basically, they allow the definition of probabilistic facts, i.e., facts that can be true or false with a certain probability. 

Following the ProbLog syntax~\cite{DBLP:conf/ijcai/RaedtKT07}, each probabilistic fact has the form $p :: a$ where $p \in ]0,1]$, meaning that $a$ is true with probability $p$ and false with probability $1 - p$. If $p = 1$, the fact is deterministic. 
An \textit{atomic choice} indicates whether a grounding $a\theta$ of a probabilistic fact $p :: a$ is selected or not, and it is represented with the triple $(a,\theta,k)$. $k \in \{0,1\}$, where $k = 1$ means that the fact is selected and $k = 0$ that it is not.
If a set of atomic choices does not contain two atomic choices in which the fact is both selected and not, it is termed \textit{consistent}.
A consistent set $\kappa$ of atomic choices is called \textit{composite choice} and its probability can be computed as:

\begin{align*}
P(\kappa) = \prod_{(a_i,\theta,1) \in \kappa} p_i \cdot \prod_{(a_i,\theta,0) \in \kappa} (1 - p_i)     
\end{align*}

A \textit{total} composite choice contains one atomic choice for every grounding of every probabilistic fact. A \textit{world} is a logic program identified by a total composite choice, and it is created by selecting the atoms corresponding to each atomic choice with $k = 1$. Its probability is given by the probability of the total composite choice. 
Here we consider only programs where every world has a two-valued Well-Founded Model (WFM). 
A semantics for programs without a two-valued WFM~\cite{well-founded} based on credal sets is discussed in~\cite{cozman2017semantics}. 
If a program does not contain function symbols, the set of groundings for every probabilistic fact is finite, and so the set of worlds for a program is finite.

\begin{example}[Card]
\label{ex:card_finite}
Consider the following toy example that models a deck of three cards composed of ace of spades, ace of clubs, and ace of hearts.
The three cards can be drawn with equal probability.
This simple scenario can be modelled using the following program:

\begin{verbatim}
1/3 :: spades(X).
1/2 :: clubs(X).
pick(0,spades) :- spades(0).
pick(0,clubs) :- \+ spades(0), clubs(0).
pick(0,hearts) :- \+ spades(0), \+ clubs(0).
\end{verbatim}





\noindent
We use only two random variables since the third (hearts) can be represented with both random variables negated (see 5th clause).
Furthermore, note that the probability of \texttt{clubs(X)} is set to $1/2$ to obtain $1/3$ for every possible choice: \texttt{P(pick(0,spades))} $= 1/3$, \texttt{P(pick(0,clubs))} $= 2/3 \cdot 1/2 = 1/3$, and \texttt{P(pick(0,hearts))} $= 2/3 \cdot 1/2 = 1/3$.
We keep the variable \texttt{X} for the first two facts, even if it may be considered a singleton variable, to better illustrate the next concepts.
\end{example}
    
Example~\ref{ex:card_finite} illustrates a program without function symbols. 
If a program has at least one variable, one constant, and one function symbol, its grounding is infinite, so the previous definition must be extended~\cite{DBLP:journals/ai/Poole97}.

The set of worlds $\omega_\kappa$ compatible with a composite choice $\kappa$ is defined as $\omega_\kappa = \{w_\sigma \in W_P \mid \kappa \subseteq \sigma\}$, where $W_P$ is the set of worlds for the whole probabilistic logic program, $\sigma$ is a total composite choice, and $w_\sigma$ is the world identified by the total choice.
If the program does not have function symbols, the probability of a composite choice is equal to the sum of the probabilities of the worlds in $\omega_\kappa$. If the program has function symbols, $\omega_\kappa$ may be uncountable and the probability of each world is 0 (infinite product of values $\in ]0,1[$). 
However, the probability of a composite choice can still be computed.
The set of worlds \textit{compatible} with a set of composite choices $K$ is given by the union of the worlds identified by each composite choice in the set. 
Two composite choices are \textit{incompatible} if their union is not consistent. 
A set of composite choices is \textit{pairwise incompatible} if every pair of different composite choices is incompatible.
$P(K)$ is still well-defined for programs with function symbols and can be computed by constructing pairwise incompatible equivalent sets.
A composite choice $\kappa$ is an \textit{explanation} for a query $q$ (a conjunction of ground atoms) if, for all worlds $w \in \omega_\kappa$, $w \models q$. 
Furthermore, a set of composite choices is \textit{covering} with respect to a query $q$ if $\forall w$ in which $q$ is true, $w \in \omega_\kappa$.
We can also define as explanation a set of worlds and define the covering property for sets of worlds.
The probability of a query $q$ is then computed by defining a probability measure over the set of worlds identified by countable sets of countable composite choices: $\mu_P(\omega_K) = \lim_{n \to \infty} \mu(K^{'}_n)$, where $K^{'}_n = \{\kappa^{'}_1, \dots,\kappa^{'}_n\}$ is a pairwise incompatible set of composite choices equivalent to $K_n = \{\kappa_1, \dots, \kappa_n\}$, $K = \{\kappa_1,\kappa_2,\dots\}$ is a covering set of explanations for $q$ and $\mu_P(K_n) = \sum_{i=1}^n P(\kappa'_i)$. Programs with function symbols have a well-defined semantics~\cite{Rig16-IJAR-IJ}.

If a probabilistic logic program is range restricted, i.e., every variable in the head appears also in a positive literal in the body, the answers of a query are always ground instantiations of it~\cite{Muggleton00,RigSwi13-TPLP-IJ}, and probabilities are always assigned to ground atoms.
However, in some cases, probabilistic facts may not be range restricted, and can contain variables (see, for instance, the probabilistic facts \texttt{spades/1} and \texttt{clubs/1} in Example~\ref{ex:card_finite}).
In this situation, queries' answers are still ground instantiations and the program can be well-defined provided that the variables in probabilistic facts are ground when the fact is called: they can be directly bound to a constant (as in Example~\ref{ex:card_finite}, where \texttt{X} is bound to \texttt{0}), or they must appear in a previous literal in the body.

\begin{theorem}[Ground queries with well-defined probability]
\label{th:th1}
Given a Probabilistic Logic Program $P$ and a query $q$, if $P$ is range restricted and all the variables in probabilistic facts appearing in the body of clauses of $P$ are present in a previous positive literal, the answers to $q$ will be ground instantiation of it, with an associated well-defined probability.
\end{theorem}
Before illustrating the proof, we introduce some basic concepts following~\cite{Muggleton00}.
Given two atoms, $a$ and $b$, a substitution $\theta$ (a replacement of variables with terms) is a unifier for $a$ and $b$ if $a\theta = b\theta$.
A substitution is a \textit{grounding} if all the involved terms are ground.
Furthermore, a unifier $\sigma$ is the most general unifier (mgu) for $a$ and $b$ if, for all unifiers $\mu$ of both $a$ and $b$, exists a substitution $\rho$ such that ($a\sigma)\rho = a \mu$.

The main step of \textit{resolution} is the following: given two clauses $A$ and $B$ with no common variables, $l \in A$ and $\neg l' \in B$, and $\theta$ being the mgu of $l$ and $l'$, the \textit{resolvent} of $A$ and $B$ is defined as $((A \setminus \{l\}) \cup (B \setminus \{l'\}))\theta$; for a program $P$ and a query $q$, resolution is linear when $B$ represents only clauses in $P$ and $A$ is the query or the resolvent of another linear resolution.
We suppose that literals in clauses are ordered and are chosen following this order during the resolution.
A resolution proof for SLDNF can be represented as a sequence of clauses $\langle q,C_1,C_2,\dots,C_n \rangle$. 
If $C_n$ is the empty clause, the proof is a \textit{refutation}. 
The substitution for the query $q$ is given by $\theta = \theta_1\theta_2\dots\theta_n$, where each $\theta_i$ is the substitution correspondent to clause $C_i$. 
Moreover, if $P$ is range restricted, $\theta$ is a grounding. 

\begin{proof}
Let us now prove Theorem~\ref{th:th1} by induction.
In the base case, the SLDNF resolution consists of only one step: the query $q$ unifies with a deterministic fact, since there is not a previous positive literal in the body that will allow the presence of a probabilistic fact.
The program is range restricted, $\theta_1$ is computed, and the variables in the query are all grounded.
Suppose now that the theorem is true at step $n$. 
The current set of substitutions is $\{\theta_1,\theta_2,\dots,\theta_n\}$.
There are two possible cases: if the selected literal of the query matches a deterministic fact, substitution $\theta_{n+1}$ grounds the literal, as deterministic facts are already ground.
If the selected literal of $q$ matches with a probabilistic fact, this fact cannot be the first of the body of the correspondent clause, by assumption: all the variables appearing in the probabilistic fact are already grounded by preceding substitutions, so the literal is ground when called, and a new substitution $\theta_{n+1}$ is computed and added to the list.
Eventually, all the variables will be grounded and the answer to the query will be a ground instantiation.
\end{proof}
If variables appear for the first time in probabilistic facts, the query could still eventually be ground, but probabilistic facts will not be ground when called, and thus the semantics will be ill-defined.

Let us now extend Example~\ref{ex:card_finite} by introducing new clauses to obtain a program with an infinite covering set of explanations.

\begin{example}[Card with an infinite number of explanations]
\label{ex:card_infinite}
Consider a game of card where a player needs to pick one out of three possible cards: ace of spades, ace of clubs or ace of hearts. 
The game stops when the player picks the ace of hearts. 
This can be modelled by adding the following clauses to Example~\ref{ex:card_finite}.

\begin{verbatim}
pick(s(X),spades):- \+ pick(X,hearts), spades(s(X)).
pick(s(X),clubs):- \+ pick(X,hearts), \+ spades(s(X)), clubs(s(X)).
pick(s(X),hearts):- \+ pick(X,hearts), \+ spades(s(X)), \+ clubs(s(X)).

at_least_once_spades :- pick(_,spades).
never_spades :- \+ at_least_once_spades.
\end{verbatim}

We can ask for the probability that the player picks at least once spades or that that he/she never picks spades (respectively \texttt{P(at\_least\_once\_spades)} and \texttt{P(never\_spades)}).

For conciseness, let us replace \texttt{spades(X)} with $f_1$ and \texttt{clubs(X)} with $f_2$. Consider query
\newline
\texttt{at\_least\_once\_spades}: it has the pairwise incompatible covering set of explanations $K=\{\kappa_0,\kappa_1,\ldots\}$ with
\begin{align*}
\kappa_0 ={}&\{(f_1,\{X/0\},1)\}\\
\kappa_1 ={}&\{(f_1,\{X/0\},0),(f_2,\{X/0\},1),(f_1,\{X/s(0)\},1)\}\\
\ldots \\
\kappa_i ={}&\{(f_1,\{X/0\},0),(f_2,\{X/0\},1),\ldots,(f_1,\{X/s^{i-1}(0)\},0),\\
&(f_2,\{X/s^{i-1}(0)\},1),(f_1,\{X/s^i(0)\},1)\}\\
\ldots
\end{align*}
In other words, in $\kappa_0$ the player picked spades at round $0$, in $\kappa_1$ he/she does not piked spades and picked clubs at round $0$ and picked spades in round $s(0)$, and so on. 

$K$ is countable and infinite and the explanations in it are pairwise incompatible, so the probability of \texttt{at\_least\_once\_spades} can be computed as
\begin{align*}
P(at\_least\_once\_spades)={}&
\frac{1}{3}+
\frac{1}{3} \cdot \left(\frac{2}{3} \cdot \frac{1}{2}\right) + \frac{1}{3} \cdot \left(\frac{2}{3}\cdot \frac{1}{2}\right)^2 + \ldots \\
={}&\frac{1}{3} + \frac{1}{3} \cdot \left(\frac{1}{3}\right) + \frac{1}{3} \cdot \left(\frac{1}{3}\right)^2 + \ldots\\
={}&\frac{1}{3} \cdot \frac{1}{1-\frac{1}{3}} = \frac{1}{3} \cdot \frac{3}{2} = \frac{1}{2}
\end{align*}
since it is the sum of a geometric series.
The probability for the query \texttt{never\_spades} can be computed in a similar way.
After the computations we get $P(at\_least\_once\_spades) = 1 - P(never\_spades) = 1/2$.



\end{example}

In the next section we introduce the syntax for hybrid probabilistic logic programs, i.e., programs with both discrete and continuous random variables.

\section{Hybrid Programs}
\label{sec:hybrid_plp}

When a probabilistic logic program contains both discrete and continuous random variables, it is called \textit{hybrid} probabilistic logic program. 
Here, we also consider constraints involving continuous random variables' values, obtaining probabilistic \textit{constraint} logic programs~\cite{AzzRigLam21-AIJ-IJ,DBLP:journals/ai/MichelsHLV15}.
In this paper we use hybrid probabilistic logic program and probabilistic constraint logic program interchangeably, since usually in PLP the goal is to compute the probability of a query (and thus we always have constraints to define the range of the continuous random variables), rather than the joint probability density induced by the program.

A probabilistic constraint logic program is composed of a set of rules, a set of Boolean probabilistic facts, and a countable set of continuous random variables.
The authors in~\cite{AzzRigLam21-AIJ-IJ} define $X = \{X_1,X_2,\dots\}$ the countable set of continuous random variables, each one associated with its range (that can be either $\mathbb{R}$ or $\mathbb{R}^n$), and $F$ the set containing discrete probabilistic facts. 
These discrete facts form a countable set of Boolean random variables $Y = \{Y_1,Y_2,\dots\}$.
The sample space of the whole program ($W_\mathcal{P}$) is defined as the product of the spaces for continuous ($W_X$) and discrete ($W_Y$) random variables, i.e, $W_\mathcal{P} = W_X \times W_Y$.

The probabilistic facts associated with every random variable $Y_i$ set to 1, and the grounding of the rules (with the constraints removed) whose constraints are satisfied given a valuation of the continuous random variables, define a ground normal logic program. 
The probability of a query can be computed by considering a pairwise incompatible covering set of explanations.
As before, an explanation for a query of a probabilistic constraint logic program is a set of worlds $\omega_i$ such that the query is true in every element of this set.

A concrete syntax for this type of programs is given by \texttt{cplint} hybrid programs~\cite{Rig18-BKaddress}.
In \texttt{cplint} hybrid programs, logical variables are partitioned into two disjoint sets: those that can assume terms as values and those that can assume continuous values. Let us call the first \textit{term} variables and the latter \textit{continuous} variables.

Continuous random variables are encoded with probabilistic facts of the form
$$A:Density$$
where $A$ is an atom with a continuous variable $Var$ as argument and $Density$ is a special atom identifying a probability density on variable $Var$. For example,
\begin{verbatim}
p(X) : gaussian(X,0,1).
\end{verbatim}
indicates that \texttt{X} in atom \texttt{p(X)} is a continuous variable that follows a Gaussian distribution with mean 0 and variance 1.
Each predicate $p/n$ has a signature that specifies which arguments hold continuous values. 
Only these arguments can contain continuous variables. 
Continuous values (and variables) can appear inside a term built on a function symbol $f/n$. 
Each function symbol $f/n$ also has a signature that specifies which arguments hold continuous values. 
Again, only these arguments can contain continuous variables.
While discrete random variables are identified by ground atoms (that form a countable set), continuous random variables are identified by predicates and by the ground terms present in atoms with arguments that can hold continuous random variables.

ProbLog probabilistic facts of the form $p::f$ can also be encoded as $f:p$ for uniformity with Logic Programs with Annotated Disjunctions~\cite{VenVer04-ICLP04-IC} and CP-Logic~\cite{DBLP:journals/tplp/VennekensDB09}.

Atoms in clauses and probabilistic facts can have both term and continuous variables. 
However, we impose the constraint that in every world of the program the values taken by term variables in a ground atom for a predicate $p/n$ that is true in the world uniquely determine the values taken by the continuous variables.

Continuous variables are introduced by probabilistic facts for continuous random variables and by the special predicate \texttt{=:=/2} that is used to define a new variable based on a formula involving existing continuous variables (see Example~\ref{ex:gm_c_ep}).
Constraints are represented by Prolog comparison predicates. 
The semantics assigns a probability of being true to any ground atom not having continuous values as arguments.
Atoms with continuous values have probability 0, as the probability that a continuous random variable takes a specific value is 0.
However, there is one special case when a continuous value is admitted as an argument. This will be discussed in Section~\ref{sec:syntax_req}.

Inference in \texttt{cplint} hybrid programs can be performed using MCINTYRE~\cite{AlbBelCot17-IA-IJ,AzzRigLamMas19-AIXIA-IC,RigBelLam16-SPE-IJ}, an algorithm based on Monte Carlo sampling.
\texttt{cplint} hybrid programs can also be translated into probabilistic constraint logic programs~\cite{DBLP:journals/ai/MichelsHLV15} by removing the continuous variables from the arguments of predicates and by replacing constraints with their probabilistic constraint logic program form.

Consider an extension to Example~\ref{ex:card_infinite}, where we also add a continuous random variable and clauses with constraints.

\begin{example}[Card extended with a continuous random variable]
\label{ex:card_continuous}
Suppose that the previous game is extended, and the player also needs to spin a wheel.
In addition to the previous rules, if the axis of the wheel is between 0 and $\pi$ degrees (approximated to $3.14$ for convenience), the game stops. In this example, we have a continuous random variable with uniform distribution that indicates the angle of the wheel, plus a constraint on its value in the clauses for the \texttt{pick/2} predicate:
\begin{verbatim}
1/3 :: spades(_).
1/2 :: clubs(_).
angle(_,X) : uniform_dens(X,0,6.28).

pick(0,spades) :- spades(0), angle(0,V), V > 3.14.
pick(0,clubs) :- \+ spades(0), clubs(0), angle(0,V), V > 3.14.
pick(0,hearts) :- \+ spades(0), \+ clubs(0), angle(0,V), V > 3.14.

pick(s(X),spades):- \+ pick(X,hearts), spades(s(X)), 
    angle(s(X),V), V > 3.14.
pick(s(X),clubs):- \+ pick(X,hearts), \+ spades(s(X)), 
    clubs(s(X)), angle(s(X),V), V > 3.14.
pick(s(X),hearts):- \+ pick(X,hearts), \+ spades(s(X)), 
    \+ clubs(s(X)), angle(s(X),V), V > 3.14.

at_least_once_spades :- pick(_,spades).
never_spades :- \+ at_least_once_spades.
\end{verbatim} 

With the query \texttt{at\_least\_once\_spades} we can compute the probability that the player picks at least one time spades when the axis of the wheel is between $\pi$ (3.14) and $2\pi$ (6.28). 
The continuous random variables are represented by the second argument of predicate $angle(T,X)$: they form a countable set, and there is a continuous random variable for each value of T (0, $s(0)$, $s(s(0))$, \dots).
The set $Y$ of discrete Boolean random variables is composed of $\{Y_i^c,Y_i^s \mid i = 1,2,\dots\}$. $Y^{c}_{i}$ ($Y^{s}_{i}$) represents $clubs(s^i(0))$ ($spades(s^i(0))$), and $y^c_i$ ($y^s_i$) are values for $Y_i^c$ ($Y_i^s$).  
Similarly, the set $X$ of continuous random variables is composed of $\{X_i \mid i = 1,2,\dots \}$.
Each $X_i$ has a range $[0,2\pi]$, and its value is denoted with $x_i$.
To compute the probability of this query, we can consider the mutually disjoint covering set of worlds $\omega = \omega_0 \cup \omega_1 \cup \dots$ where

\begin{align*}
\omega_0 ={}&\{(w_{\mathrm{X}},w_{\mathrm{Y}}) \mid w_{\mathrm{X}} = (x_1,x_2,\ldots), w_{\mathrm{Y}} = (y^{c}_1,y^{s}_1,y^{c}_2,y^{s}_2,\ldots), \\ 
&x_1 \in [\pi,2\pi],y^s_1 = 1\}\\
\omega_1 ={}&\{(w_{\mathrm{X}},w_{\mathrm{Y}}) \mid w_{\mathrm{X}} = (x_1,x_2,\ldots), w_{\mathrm{Y}} = (y^{c}_1,y^{s}_1,y^{c}_2,y^{s}_2,\ldots), \\ 
&x_1 \in [\pi,2\pi],y^s_1 = 0,y^c_1 = 1, x_2 \in [\pi,2\pi],y^s_1 = 1 \}\\
\ldots \\
\end{align*}
That is, for explanation $\omega_0$ spades was selected at round 0 ($y^s_1 = 1$) and the wheel ($x_1$) in the same round was in the range $[\pi,2\pi]$.
For explanation $\omega_1$, spades was not selected at round 0 ($y^s_1 = 0$), clubs was selected at round 0 ($y^c_1 = 0$), the wheel ($x_1$) was in the range $[\pi,2\pi]$ at round 0, spades was selected at round $s(0)$ ($y^s_2 = 1$) and the wheel ($x_2$) was in the range $[\pi,2\pi]$ at round $s(0)$.
The probability for $\omega_0$ can be computed as~\cite{AzzRigLam21-AIJ-IJ,chow2012probability}:

\begin{align*}
\mu(\omega_0) ={}&\int_{\pi}^{2\pi} \mu_{\mathrm{Y}}(\{(y^{c}_1,y^{s}_1,y^{c}_2,y^{s}_2,\ldots) \mid y^c_1 = 1\}) \ d\mu_{\mathrm{X}} \\ 
={}&\int_{\pi}^{2\pi} \frac{1}{3} \cdot \frac{1}{2\pi} dx_1 = \frac{1}{3} \cdot \frac{1}{2} = \frac{1}{6}.
\end{align*}
where $\frac{1}{3}$ is the contribution of the discrete random variable (spades) and $\frac{1}{2\pi}$ is the contribution of the continuous one (angle). 
The probability for the other $\omega_i$ can be computed in a similar way.
Overall, considering the limits, we get $\frac{1}{3} \cdot \frac{1}{2} \cdot \sum_{i=0}^{\infty} (\frac{2}{3} \cdot \frac{1}{2} \cdot \frac{1}{2})^i = \frac{1}{6} \cdot \sum_{i=0}^{\infty} (\frac{1}{6})^i = \frac{1}{6} \cdot \frac{6}{5} = \frac{1}{5}$ as probability for the query \texttt{at\_least\_once\_spades}.


\end{example}

\begin{example}[Gaussian mixture]
\label{ex:gm_cplint}
A Gaussian mixture model is a way to generate values of a continuous random variable: a discrete random variable is sampled and, depending on the sampled value, a different Gaussian distribution is selected for sampling the value of the continuous variable.

A Gaussian mixture model with two components can be expressed in \texttt{cplint} hybrid programs as~\footnote{\url{http://cplint.eu/e/gaussian_mixture.pl}}:

\begin{verbatim}
h : 0.6.
heads :- h.
tails :- \+ h.
g(X) : gaussian(X, 0, 1).
h(X) : gaussian(X, 5, 2).
mix(X) :- heads, g(X).
mix(X) :- tails, h(X).
mix :- mix(X), X > 2.       
\end{verbatim}

\noindent
The argument \texttt{X} of \texttt{mix(X)} follows a distribution that is a mixture of two Gaussians, one with mean 0 and variance 1 with probability 0.6, and one with mean 5 and variance 2 with probability 1 - 0.6 = 0.4.
We can then ask for the probability of \texttt{mix}.

Here, predicates \texttt{g/1}, \texttt{h/1}, and \texttt{mix/1} have a single argument which can hold continuous variables, so overall there is a finite set of continuous random variables. 
Since there are no term variables, each atom for these predicates in a world univocally determines its argument. 
For predicate \texttt{mix/1} this is not obvious as there are two clauses for it.
However, the two clauses have mutually exclusive bodies, i.e., in each world only one of them is true. This property is further discussed in Section~\ref{sec:syntax_req}.
\end{example}

\begin{example}[Gaussian mixture and constraints, from~\cite{TLP:8688161}]
\label{ex:gm_c_ep}
Consider a factory with two machines, \texttt{a} and \texttt{b}. Each machine produces a widget with a continuous feature.
A widget is produced by machine \texttt{a} with probability 0.3 and by machine \texttt{b} with probability 0.7.
If the widget is produced by machine \texttt{a}, the feature is distributed as a Gaussian with mean 2 and variance 1.
If the widget is produced by machine \texttt{b}, the feature is distributed as a Gaussian with mean 3 and variance 1.
The widget then is processed by a third machine that adds a random quantity to the feature. 
The quantity is distributed as a Gaussian with mean 0.5 and variance 1.5.
This can be encoded by in \textup{\texttt{cplint}} hybrid programs as~\footnote{\url{http://cplint.eu/e/widget.pl}}:

\begin{verbatim}
machine(a) : 0.3.
machine(b) :- \+ machine(a).
st(a,Z) : gaussian(Z, 2, 1).
st(b,Z) : gaussian(Z, 3, 1).
pt(Y) : gaussian(Y, 0.5, 1.5).
widget(X) :- machine(M), st(M,Z), pt(Y), X =:= Y + Z.
ok_widget :- widget(X), X > 1.0.    
\end{verbatim}

\noindent We can then ask the probability of \texttt{ok\_widget}.

Here, \texttt{X}, \texttt{Y}, and \texttt{Z} are continuous variables and \texttt{M} is a term variable. 
Since \texttt{X} is a continuous variable, in every world there should be a single value for \texttt{X} that makes \texttt{widget(X)} true. 
Predicate \texttt{widget/1} has a single clause but the clause has two groundings, one for \texttt{M} = \texttt{a}, and one for \texttt{M} = \texttt{b}, so in principle there could be two values for \texttt{X} in true groundings of \texttt{widget(X)}. 
However, as in Example~\ref{ex:gm_cplint}, the two groundings of the rule have mutually exclusive bodies, as in each world either \texttt{machine(a)} is true or \texttt{machine(b)} is true, but not both.
\end{example}

\begin{example}[Estimation of the mean of a Gaussian]
\label{ex:mest_cplint}
Consider now this example~\footnote{\url{http://cplint.eu/example/inference/gauss_mean_est.pl}}:

\begin{verbatim}
mean(M) : gaussian(M,1,5).
value(_,M,X) : gaussian(X,M,2).
value(I,X) :- mean(M), value(I,M,X).    
\end{verbatim}

The program states that, for an index \texttt{I}, the continuous variable \texttt{X} is sampled from a Gaussian whose variance is 2 and whose mean \texttt{M} is sampled from a Gaussian with mean 1 and variance 5.

This program can be used to estimate the mean of a Gaussian by querying \texttt{mean(M)} given observations for atom \texttt{value(I,X)} for different values of \texttt{I}.

Here, the first argument of \texttt{value/3} can hold a term variable while its second and third arguments can hold a continuous variable.
The second argument is used as a parameter in the probability density of the third argument.
It is not immediate to see whether this program has a well-defined semantics or not. 
In fact, the semantics does not allow specifying the parameters of continuous distributions with values computed by the program, but we can consider continuous variables \texttt{M} and \texttt{X} as specified by a joint density (see Section~\ref{sec:syntax_req}).
Since a Gaussian density with a Gaussian mean is still a Gaussian, the joint density will be a multivariate Gaussian.
\end{example}

In the next Section, we illustrate more in detail the syntactic requirements that hybrid probabilistic logic programs, and thus probabilistic constraint logic programs, must follow to have a well-defined semantics.

\section{Syntactic Requirements}
\label{sec:syntax_req}
To preserve the well-definedness, we require that the set of random variables must be countable.
That is, every random variable must be associated with a ground logical atom.
The discrete arguments of this logical atom can contain terms and cannot be real values.
Let us now analyse several possible situations.

In the fourth clause of Example~\ref{ex:card_continuous}, (reported here for clarity)

\begin{verbatim}
pick(0,spades) :- spades(0), angle(0,V), V > 3.14.
\end{verbatim}
the first argument of the continuous random variable \texttt{angle} is a constant. 
Similarly happens with the clauses with \texttt{angle(s(X),V)}.
Variable \texttt{X} appears in the head, so it is ground when \texttt{angle(s(X),V)} is called (during the recursion or by directly set a value for \texttt{X} in the query), and the continuous random variable \texttt{V} is associated with term \texttt{s(X)} by predicate \texttt{angle/2}.
In both cases, \texttt{s(X)} is a ground logical term (integer in case of \texttt{0}), so the semantics is well-defined. 

Consider now Example~\ref{ex:gm_cplint}: in this program there are two random variables, identified respectively by \texttt{g(X)} and \texttt{h(X)}. 
Predicate \texttt{mix/1} is composed of two clauses with the same head but different bodies, a situation that may lead to an ill-defined program where both clauses are true and the random variable \texttt{X} is defined by two different distributions.
However, the two bodies are mutually exclusive: the discrete random variable \texttt{h} is used to discriminate between the two.
When \texttt{h} is true, \texttt{heads} is true, and thus the first clause of \texttt{mix/1} is considered. 
When it is false, the second \texttt{mix/1} clause is considered.
This mutual exclusivity guarantees the well-definedness of the program.
Algorithmically, this property can be verified performing clause unfolding, obtaining two clauses with the same head with bodies mutually exclusive, since they contain at least one atom in common, but in one it is positive, in the other it is negated.
The same happens for Example~\ref{ex:gm_c_ep}.
This idea can be extended to the case where there are $n$ clauses with the same head defining $n$ different distributions for the same variable, provided that all the clauses are mutually exclusive.

To see a counterexample, if we modify the clauses for predicate \texttt{mix/1} of Example~\ref{ex:gm_cplint} in this way:
\begin{verbatim}
mix(X) :- g(X).
mix(X) :- h(X).
\end{verbatim}
the program is ill-defined, since the two clauses are not mutually exclusive, and there is not a single distribution for variable \texttt{X}.

Focus now on Example~\ref{ex:mest_cplint}. 
Here, atom \texttt{value/3} has a continuous random variable (\texttt{M}) as input. 
To keep the well-definedness, this input must be defined in another continuous probabilistic fact and must be used only as a parameter for another distribution.
In fact, \texttt{M} is obtained from a Gaussian distribution and it is used as a parameter for the mean for another Gaussian distribution.
In this way, variables \texttt{M} and \texttt{X} are specified by a joint density. 
Note that the first variable of \texttt{value/3} is a term variable and can be associated with a ground atom (integer, for example), preserving the well-definedness: the value of the term variable uniquely identifies the value of the continuous random variable.
In this way, the position in the sequence of random variables is exactly identified.

This situation can be straightforwardly extended to programs in which there are multiple continuous random variables as input: for instance, in Example~\ref{ex:mest_cplint}, the variance for \texttt{X} in clause \texttt{value/3} can be sampled from another distribution instead of being fixed to 2.

In other cases, when the value of a continuous random variable is used as variable for another term, but not as a parameter for a distribution, the semantics is ill-defined. 
Consider the following modification to Example~\ref{ex:mest_cplint}:

\begin{verbatim}
value(X) : gaussian(X,1,5).
value(_,M) : gaussian(M,2,2).
res(M) :- value(X), value(X,M). 
\end{verbatim}

Here, \texttt{X} in clause \texttt{res/1} is the value of a continuous random variable but it is not used as parameter for the distribution of \texttt{M}.
The continuous random variable \texttt{M} cannot be associated with a term that uniquely identifies it, so the program is ill-defined. 
The difference with Example~\ref{ex:mest_cplint} is that in Example~\ref{ex:mest_cplint} the term variable used as index (\texttt{I}) can be associated with a ground logical term, while here it is associated with a real value (\texttt{X}).
If in this last example the variable \texttt{X} is a discrete random variable rather than a continuous one, the program would be well-defined, since \texttt{X} is associated with a ground logical atom.

To see how to compute the probability in this situation, consider the following simple program (a simplification of Example~\ref{ex:mest_cplint}, where Gaussian distribution are replaced by Uniform distributions identified by \texttt{uniform\_dens/3}):

\begin{example}
\label{ex:float_param}
\begin{verbatim}
angle_a(_,X,Y) : uniform_dens(Y,X,2).
angle_b(X) : uniform_dens(X,0,1).
success(I) :- angle_b(X), angle_a(I,X,Y), Y < 1.5.
\end{verbatim}
\end{example}
Here, the variable \texttt{X} in \texttt{angle\_b/1} follows a uniform distribution between 0 and 1. 
For variable \texttt{Y} in \texttt{angle\_a/1}, its lower bound is sampled from another uniform distribution.
The probability distribution defined by the two random variables can be considered as a joint distribution between the two, so they can be treated as one multivariate random variable in the sequence of random variables, indexed by the logical term unified with \texttt{I}.
Their joint density function is given by
$$
f_{XY} (x,y) = \frac{1}{1 - 0} \cdot \frac{1}{2 - x} = \frac{1}{2-x}
$$

and thus the probability of the query \texttt{success(0)} is

\begin{align*}
P(success(0)) = P[Y < 1.5] &= \int_{0}^{1} \frac{1}{2 - x} \left( \int_{x}^{1.5} 1\ dy \right) dx = \\
&= \int_{0}^{1} \frac{1.5 - x}{2 - x} \ dx = 1 - \frac{ln(2)}{2} \approx 0.653  
\end{align*}

As said in Section~\ref{sec:plp}, the range-restrictedness property ensures that the answers of a query are always ground instantiations of it.
However, there can be some exceptions.
Consider the first three lines of Example~\ref{ex:card_continuous}, reported here for clarity:
\begin{verbatim}
1/3 :: spades(_).
1/2 :: clubs(_).
angle(_,X) : uniform_dens(X,0,6.28).
\end{verbatim}

Both discrete and continuous probabilistic facts are not range restricted, since they contain an input variable (anonymous) that is not ground.
To ensure that the answers of a query are always ground instantiations of it, we need to make sure that, when these types of facts are called, all the input variables are ground.
We then impose that the arguments of probabilistic facts that are not ground must be variables that appear in previous literal in the body: the preceding calls will bind these variables, so the probabilistic facts will be called with all input arguments ground, and a probability value will be assigned to a ground query.
The order of the terms is fundamental: the variables in a probabilistic fact must appear in a preceding literal, to ensure the well-definedness.
Consider the following clarifying example:
\begin{verbatim}
a(1).
f(_,X) : uniform_dens(X,0,6).

g0(X):- a(X), f(X,V), V > 1.
g1(X):- f(X,V), V > 1, a(X).
\end{verbatim}
For the query \texttt{g0(X)}, when \texttt{f(X,Y)} is called, \texttt{X} has already been substituted with \texttt{1}, and \texttt{f/2} has a ground input variable.
This does not hold for query \texttt{g1(X)} (even if the terms in the body are the same, they are in different order), since \texttt{f/2} is called with the input variable not instantiated, thus violating the syntactic requirements.

To sum up, if we consider the following simple program:
\begin{verbatim}
ud(_,V) : uniform_dens(V,0,6).
1/2 :: a(_).
b(X,1):- a(X).
b(X,2):- \+a(X).
n(1).
f_0(X):- a(X).
f_1:- a(1).
f_2(X):- n(X), a(X).
f_3:- a(_).
f_4(X):- b(X,V), a(V),
f_5:- ud(1,V), V > 2.
\end{verbatim}
the answers of the queries \texttt{f\_0(1), f\_1, f\_2(X)}, and \texttt{f\_5} are ground instantiations of them, since the input term variable for probabilistic facts \texttt{a/1} and \texttt{ud/2} is bounded to \texttt{1} in all the queries.
Similarly happens for \texttt{f\_4(1)}, where \texttt{V} can unify with \texttt{1} or \texttt{2}, depending on the truth value of \texttt{a}.
On the contrary, the answers for \texttt{f\_0(\_), f\_3}, and \texttt{f\_4(\_)} are not ground instantiations, since the term variable of the probabilistic fact \texttt{a} is not ground (for the first two queries) or the input variable \texttt{X} of \texttt{b/2} is not ground (for the third query).

\section{Conclusions}
\label{sec:conclusions}
In this paper, following the semantics proposed in~\cite{AzzRigLam21-AIJ-IJ}, we delineated a precise syntax and explained the necessary syntactic conditions to maintain the well-definedness for hybrid probabilistic logic programs.
We covered several cases in which ill-definedness may arise, providing examples and counterexamples.
As future work, we plan to develop algorithms to automatically check these properties as well as developing new inference algorithms that can manage infinite domains~\cite{10.1007/978-3-030-58449-8_1}.

\bibliographystyle{eptcs}
\bibliography{bibtexrepository/journals_long,bibtexrepository/booktitles_long,bibtexrepository/publishers_short,bibtexrepository/series_springer,bibtexrepository/publishers_long,bibtexrepository/bibl}

\end{document}